# Online Rating System Development using Blockchain-based Distributed Ledger Technology

Monir Shaker[1] . Fereidoon Shams Aliee[1] . Reza Fotohi[1]

**Abstract** In most websites, the online rating system provides the ratings of products and services to users. Lack of trust in data integrity and its manipulation has hindered fulfilling user satisfaction. Since existing online rating systems deal with a central server, all rating data is stored on the central server. Therefore, all rating data can be removed, modified, and manipulated by the system manager to change the ratings in favor of the service or product provider. In this paper, an online rating system using distributed ledger technologies has been presented as the proposed system to solve all the weaknesses of current systems. Distributed ledger technologies are completely decentralized and there is no centralization on them by any institution. Distributed ledger technologies have different variants. Among distributed ledger technologies, blockchain technology has been used in the proposed rating system because of its support for smart contracts. In the proposed online rating system, the Ethereum platform has been chosen from different blockchain platforms that have a public permission network. In this system, the raters cannot rate unless they submit a request to the system and be authorized to take part in the online product rating process. The important feature of the Ethereum platform is its support for smart contracts, which can be used to write the rating contract in the Solidity language. Also, using Proof of Authority (PoA) consensus mechanisms, all rating transactions are approved by the surveyors. Since in the real Ethereum system, each rating transaction is sent to the network by the raters, some gas must be paid for each rating transaction. However, since this method is expensive, TestNet blockchain can be used in the rating system. Finally, the proposed rating system was used for rating the restaurants of a website and its features were tested.

**Keywords:** Online Rating System, Distributed Ledger Technology, Blockchain, Proof of Authority (PoA); TestNe

✉ Monir Shaker
  moneershaker@gmail.com

✉ Fereidoon Shams Aliee
  f_shams@sbu.ac.ir

✉ Reza Fotohi*
  R_fotohi@sbu.ac.ir; Fotohi.reza@gmail.com

[1] Faculty of Computer Science and Engineering, Shahid Beheshti University, G. C. Evin, Tehran, Iran

# 1 Introduction

Digital transformation is the integration of digital technology at all business levels and it fundamentally concentrates on changing how operations are carried out and delivering value to the customers. In addition to that, it addresses the organizational culture modifications needed for dealing with the challenges of the current situation. Digital transformation is a significant change in the operation of an organization or a country with a focus on transformational technologies. Internet of Things, cloud computing, mobile applications, social media, virtual and augmented reality, data analysis, artificial intelligence, and blockchain are some of the most important transformational technologies [1]. Blockchain technology is among the digital transformation technologies and it is a Distributed Ledger Technology (DLT). It is a fully autonomous and independent network that, due to its decentralized nature, the chance of infiltration and malicious activity is very low in this network. The main advantages of blockchain technology are due to the lack of a central core. The combination of the transparency of distributed ledger and the security of encrypting an unchangeable data stream has made this technology an ideal tool for the interactions between businesses and validating information between them. Blockchain technology is a fundamental technology that can be configured in various ways according to different goals and business models [2].

In the modern world, rating is an applied element and is used in many organizations and institutions. Until now, the rating has mainly been done offline and had numerous problems and limitations. The offline system is very expensive, time-consuming, and usually inaccessible and it is considered a big challenge for those who cannot be physically present for rating. For this reason, it seems like online rating systems can be a suitable alternative for the offline system because the online system is capable of eliminating, or at least helping with the elimination of, the challenges and the problems present in the offline system. The online rating system is one of the electronic city services and it is very important in the fate of a country. Currently, the security requirements are being studied. The online rating system is an electronic solution on the way to technological advancement. All government institutions and organization managers emphasized that such a rating system is needed to create efficiency in the business. The important matter here is that using the online system has not been satisfactory till now and many users of organizations and departments who used the online system, do not trust the correctness of the ratings of this system enough. The reason that has caused online rating to have some weaknesses, including not being able to fulfill the satisfaction of users and organizations, is that the concern about data integrity is most pronounced in the online system. In other words, the lack of trust in the integrity of data and its manipulation has hindered fulfilling user satisfaction. For instance, Yelp is an application that helps people find their intended restaurants and cafés. In addition to that, this application rates the restaurants and cafés using users' opinions. Therefore, customers and users go to their desired restaurants and cafés based on the results provided by Yelp [3].

However, the problem is that many Yelp users do not completely trust their ratings and often think that these ratings might have been manipulated in Yelp. To solve this problem, in this research, it has been tried to solve the issue of lack of trust in the ratings of such websites by using distributed ledger technologies.

One of the methods used in strategic planning for organizations is the SWOT matrix. This matrix includes the threats, opportunities, weaknesses, and strengths of a system. Figure 1 presents the SWOT matrix for the online rating system. As shown in the figure, the weaknesses of this system include the lack of transparency, untrustworthiness, and lack of privacy. In this research, we try to solve the lack of transparency and untrustworthiness weaknesses of the system by using distributed ledger technologies.

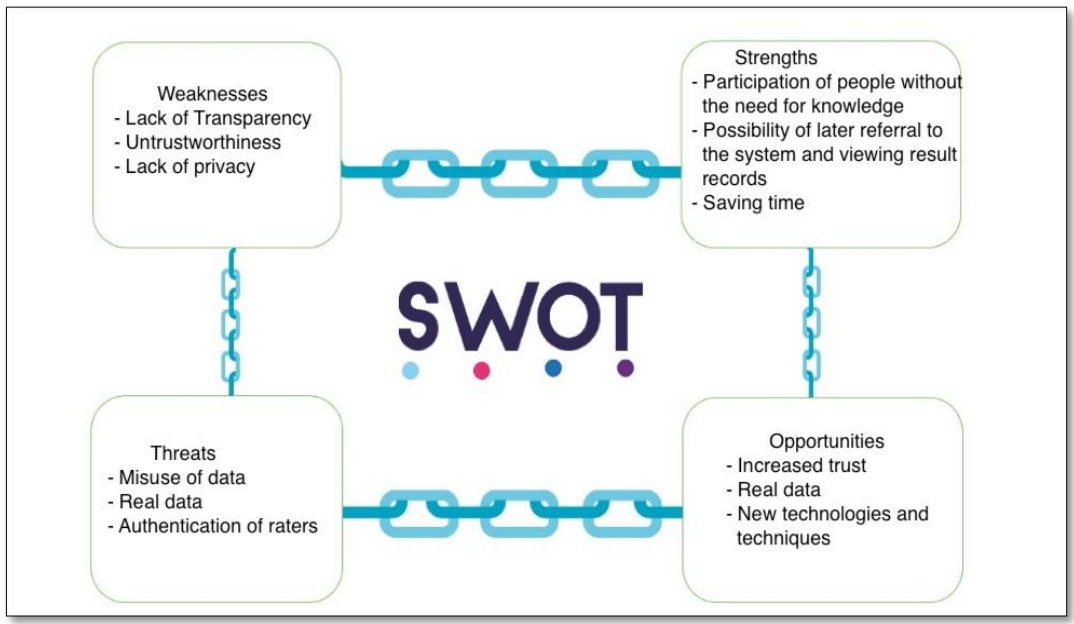

**Fig. 1.** SWOT matrix of the online rating system

Considering the features of distributed ledgers, in this research, it has been tried to use the advantages of this technology for eliminating the problems of the rating system. Therefore, the features of different distributed ledger technologies are studied to find out which one is appropriate for use in the rating system. Afterward, this technology is used to make the rating process nonmanipulable and trustworthy. Also, it is tried for the rating to only be done online and by individuals (raters) or users and also for no person to have the ability to make changes to the rating results by manipulating information. Finally, the matter of users being able to make sure their ratings and comments are recorded in the system after they are done rating has been considered.

The paper presented here is organized as the following. In Section 2, fundamental concepts, which are used in this paper, are presented. In Section 3, we will review the related work. Section 4 provides our proposed online rating system with blockchain. The parameters used for assessing the performance are studied, and simulation outcomes are deliberated in Sect. 5. Section 6, we present the results obtained during this research and provide some suggestions for future work.

## 2 Preliminaries

In this section, the fundamental concepts used in this paper, including digital transformation, distributed ledger technologies, smart contracts, online rating system with blockchain, comparison of voting and rating systems, role of blockchain technology in key goals of rating systems, and blockchain technology challenges are discussed.

### 2.1 Digital Transformation

Digital transformation is the integration of digital technology in all business fields and it fundamentally concentrates on changing the way business operations are carried out and delivering value to the customers. Furthermore, it focuses on making changes to the organizational culture required for coping with the challenges of the current condition [1]. Becoming a digital business means using technology for creating new types of products and procedures instead of existing ones. Digital strategies depend on the application of digital properties in the new methods. These properties include analyst software, big data, cloud, internet of things, developing applications, and blockchain. Digital transformation is not separate from technology. However, it requires cultural elements that encourage the company to immediately change with an ever-changing business prospect. Of course, This is not easy. The digital world lets companies provide the customers' needs through new methods. Digital transformation is the usage of digital technologies for making gradual fundamental changes in the principles and rationale of businesses and industries. As presented in figure 2, digital technologies include the internet of things, robotic process automation, cloud computing, blockchain, artificial intelligence, and many other new technologies.

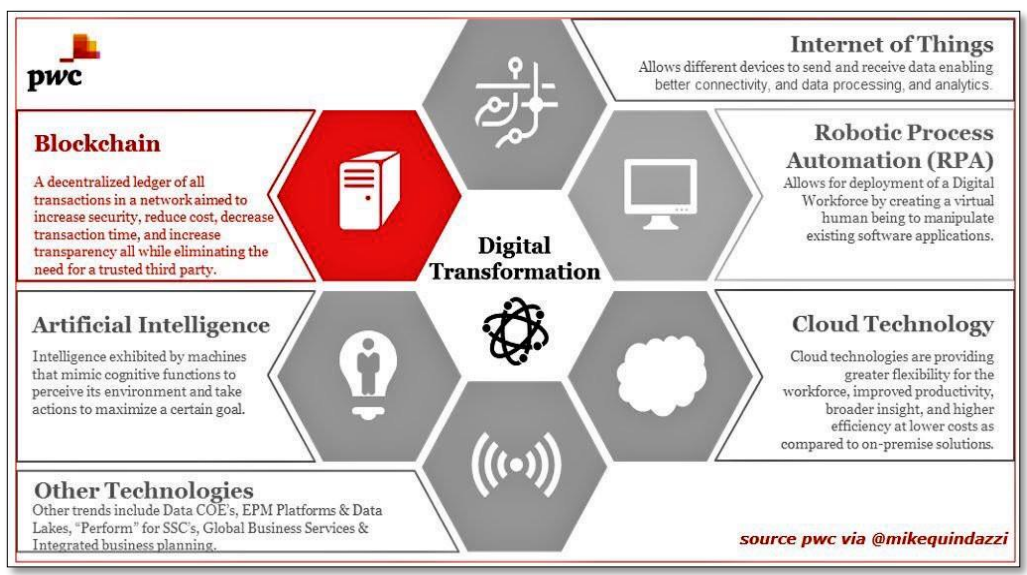

**Fig. 2.** Digital transformation technologies [4].

## 2.2 Distributed Ledger Technologies

The current conventional method for storing data is to use a database situated in a central server. For instance, all information on bank transactions is located in a central database on a server that belongs to the bank. If this database is attacked, maybe the bank has a backup database that can recover the intended information. But if the backup version is also attacked, what can be done? In that case, the organization should have several backups to ensure the safety of its information. In figure 3, the centralized part shows this type of network.

Distributed ledger is a database which is completely shared and syncs across the whole network and is distributed among several sites, organizations, and locations. Distributed ledger lets the data have public witnesses and therefore makes cyber-attacks more difficult [5]. Participants in each network node can access the record shared on that network and they have access to a similar version of the records. Furthermore, any change, modification, or addition to the ledger is versioned for all participants and applied in a matter of seconds or minutes. In figure 3, the move from a centralized network to distributed/decentralized ledger is presented:

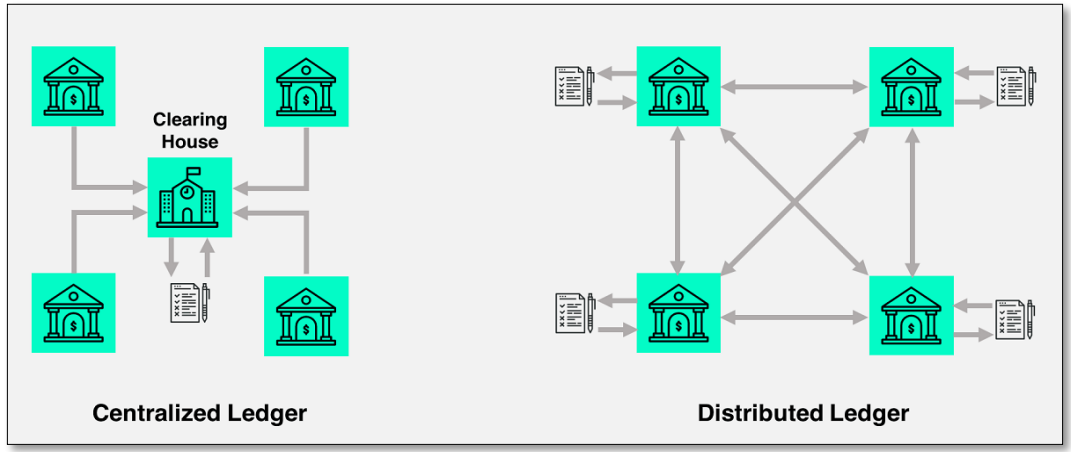

**Fig. 3.** From a centralized ledger to distributed ledger [2].

Distributed ledger technology solves the problem of attacking the centralized system. Blockchain is a type of distributed ledger that has its specific mechanisms. In this ledger, information and data is encrypted and saved in interconnected blocks [6].

### 2.3 Smart Contract

An ordinary contract is an agreement between two or more people that commits them to something in the future. The thing that leads to the difference between ordinary and smart contracts is the computer codes that resolve the need for trust. As an example, Ali creates a contract with an insurance company, according to which he pays a sum of money to the insurance company and the company guarantees to pay the possible damages to Ali's car in the following year [4].

Now, if the contract is a smart contract, the codes automatically check the circumstances. For instance, has Ali paid his insurance fee? Has Ali's car damage been submitted? And a decision is made automatically if any of the conditions are met, for example, the contract is cancelled or renewed.

The main feature of a smart contract is its execution without the need for trust, that eliminates the need for a third party for executing the different circumstances of the contract. Instead of relying on a person, who can make mistakes intentionally or unintentionally, at the appropriate time it accurately executes what it has been programmed to do.

At first, all the assets and contract conditions are encoded and stored in the blockchain. This contract is distributed and copied several times among the platform nodes. After processing is done, the contract is executed according to specified circumstances. In smart contracts, an individual or institution cannot control a contract and when the contents of a contract are correct, it is executed completely automatically.

The following items are needed for creating a smart contract [7, 8]:

1. Contract subject: this program must have access to the services or products in the contract to automatically lock or unlock them.
2. Digital signature: all participants will sign the agreement by signing the contract using their private keys.

3. Contract terms: terms of smart contracts are in the form of a precise sequence of operations that must be done. All participants must accept these terms.
4. Platform: smart contracts are developed on the blockchain platform and distributed among the nodes of this platform. Each smart contract has several parts including address, different variables, and input and output functions.

**Address:** when the smart contract is executed, it gets a unique address. This address is saved in the blockchain.

**Variables:** these are used in smart contract functions. These variables are stored in validators' data. When sent, each transaction contains the contract address and the name of the function that should be called along with the variables of that function. After the transaction is stored in the blockchain, it is visible to all validators

## 2.4 Online rating system with blockchain

Currently, users of websites providing products or services can get information about the product or service before making a purchase by reviewing other users' ratings and make an informed decision. There are some flaws in the current online rating system. For instance, users are not sure of the accuracy of information (ratings) and it is possible for the providers to manipulate the ratings and change the results in their favour using various methods. However, decentralized online voting using blockchain technology can create and verify a unique digital identity for the users. This way, all the data is encrypted and saved in the blockchain to revive trust and transparency in online markets and enable the users to trust different online platforms.

Online sellers and voters can easily create multiple user accounts for each system to improve the ratings, which the decentralized system restricts. Fake ratings are booming these days and sellers and merchants are affected by this. It is currently possible to buy positive ratings for your own benefit or to buy negative ratings for affecting your competitors to increase online credit. This causes users to have no trust in the ratings of different platforms or even websites. Nevertheless, using blockchain technology is a way to change the management of online rating. The thing that has hindered online rating from satisfying the users and organizations is that in the online system, the biggest concern is regarding data accuracy and manipulation.

A decentralized online rating system becomes trustworthy using smart contracts and it helps the ratings to be submitted by individuals who have been verified in the system. Therefore, no one can ever make any changes in rating results by manipulating data.

## 2.5 Comparison of voting and rating systems

In the following table, the similarities and differences of voting and rating systems are expressed

**Table 1** Comparison of voting and rating systems.

|  | Rating system | Voting system |
| --- | --- | --- |
| Definition | Collecting numeral ratings about products or services | Decision process for choosing people |
| When can opinions be expressed? | Unlimited time Registering ratings after the creation of product | After the agenda is prepared and usually for a limited amount of time |
| How is polling done? | Facility in rating by seeing the created product | Filling the form |
| Where are the opinions stored? | Blockchain technology ||
| How are the opinions verified? | Information is validated using consensus algorithms ||

As a result, the voting system should reach the end of the voting process to declare the votes but the rating process does not have this property. The number of ratings must be declared each second in the rating system.

### 2.6 Role of blockchain technology in key goals of rating systems

In the online rating system, blockchain technology plays an important role in different aspects of the rating system. Table 2 presents the role of blockchain technology on features like transparency, trust, preventing the manipulation of data, and privacy, which are some of the features of the rating system.

**Table 2** Role of blockchain technology in the online rating system.

| Aspects of the rating system | Utilized solutions | Role of blockchain |
| --- | --- | --- |
| Transparency and trust | Smart contract | Trust between the parties providing services or products and rater in the rating system |
|  | The digital signature can validate the identity of people and ratings | Blockchain-based digital license as a tool for increasing trust |
| Reducing danger and preventing rating data manipulation | Authentication of individuals participating in the rating process | Only parties admitted to the network are involved in rating and other situations |
| Privacy | Encrypted public and private keys | Increasing the confidentiality of each rater |

### 2.7 Blockchain technology challenges

In this section, we will discuss some of the possible challenges of blockchain. These challenges include:

**Limited scalability:** Blockchain is a peer to peer system that follows two things: on the one hand, blockchain allows anyone to add new transaction data to the collectively-stored records, and on the other, it ensures the data record is immune to manipulation and forgery. Blockchain beautifully achieves both goals using an additive-only and unalterable data structure. Such that whenever a new block is added, it requires solving a hash puzzle. Solving this hash puzzle has intentionally been designed to be time-consuming. Insisting on solving the hash puzzle is a suitable way of making any effort to manipulate transaction data records very expensive. Unfortunately, this security measure leads to trade-offs like decreased processing speed and therefore limited scalability [9].

**High cost:** The matter of high cost relates to the issue of limited scalability. Solving the hash puzzle or providing proof of work license is computationally expensive. All these actions are for security that leads to the invariability of transaction data records. The computational cost can be expressed in different scales like the number of computational cycles, time spent, and money [10].

**Validators' strategies:** It has been proven that validators can earn more than their fair share using a selfish mining strategy. Validators hide their blocks for more income in the future. In this method, branches can happen constantly. This hinders the development of blockchain [10].

## 3 Related work

In this section, we explain some of the most important blockchain-based online rating and grading systems that provide blockchain-based online elections.

The problem with the academic debate is that a number of reputable academic publishers today are dominated by a number of key figures in the field, and as a result of this imbalance in the dissemination of academic work, the whole process becomes inefficient and useless. In this paper, a scientific publishing platform based on China Blockchain technology is proposed to be developed to solve the problem of imbalance in the dissemination of academic work. This article provides a fair distribution of rewards for all shareholders and immediate ownership to the authors of an article. The Ethereum platform is used and this platform includes backend and frontend. Interactive processes of users who are writers and judges in this process. Therefore, in this new system, a page is shown for the authors to submit the article and from the judges who can review the articles to show positive or negative approval, and the whole process is performed in the Ethereum blockchain system [11].

With the development of communication technology in the field of tourism, tourists share information related to the ranking of travel destinations through their mobile devices. The most common method of sharing the data used is the centralized model. However, centralized architecture has several weaknesses, so it should be used through the decentralized architecture developed in this research in a tourist destination ranking system (TDRS). This paper presents a hybrid system of 6AsTD as a framework for assessing the level of success of tourism destinations and China Blockchain technology security system as a data sharing architecture between tourists. AsTD Framework 6 uses six variables as a reference for evaluating tourism destinations. The ranking results were distributed to each mobile device connected to China Blockchain. In this article, it was tested on several tourists [12].

The current e-commerce business model is not reliable for different customers and is questionable and every customer is looking for a reliable supplier. Suppliers with different product quality may even have customers of reputable e-commerce companies, for example Amazon and Alibaba do not adopt quality grades to evaluate products. In this paper, we propose a blockchain technology (BPGS) system to deal with big data for these business models. Blockchain is a decentralized technology, so we can expedite product grading. In addition, for the proposed BPGS system, 51% of attacks cannot be performed unless 51% of e-commerce suppliers and companies are simultaneously at risk [13].

A peer-to-peer energy system allows participants to exchange energy directly with each other. A pricing mechanism can regulate the price of trade between merchants, and a scoring metric can encourage participants to behave well. This paper proposes a peer-to-peer energy trading model, including a blockchain framework to guarantee merchant information and execution, a two-tier pricing mechanism to determine the transaction price for each transaction, and a credit rating system to improve market quality such as For example. This pricing mechanism allows manufacturers to achieve maximum cost or revenue savings, and the credit rating system encourages them to improve their reputation so that they have an advantage over other competitors during trading. . Therefore, the design of these methods to achieve the relevant tasks and create a closed chain is discussed in this article. As a result, a case study shows the effect of the proposed peer-to-peer business model [14].

Ring economics will be a key concept in government promotion, emphasizing reconstruction over resource ownership, and proposing the use of shared resources to create new supply chains and new economies. When implementing a cyclical economy, each economic entity must learn how to rank each other, and this process is time consuming. This paper proposes a cyclical credit rating system based on the Chinese blockchain. It uses China Blockchain technology to detail the transactions of each entity, and then uses trust-level algorithms to calculate each entity's credit rating. This method uses the concept of decentralization to reduce third-party intermediary costs, so that, in addition to reducing transaction costs, it provides an effective credit rating of public economic entities, thus establishing relationships between the two parties at all times and establishing their transaction [15].

Traditional small lending is a new online database where people from all over the world can lend their money to those who need it. The problem with this traditional method is that the risk of non-payment is very high compared to traditional financing, which is why the platform needs to be transparent and reliable. Therefore, in this paper, they propose a model that combines hyper location fabric blockchain technology with a rating system to validate both the borrower and the lender, rather than the institutions and companies that act as intermediaries with a peer-to-peer online platform. Peers are in touch. This proposed model provides security with system openness and can be integrated into existing systems for a more efficient system [16].

Communication between IoT tools is not secure and reliable. There are a number of security risks involved. Existing security mechanisms are not easy to manage because they require additional resources and thus increase the overall cost of the system. IoT devices are limited resource tools and do not perform many calculations. Services requested from IoT devices may be malicious and when not intercepted by any effective security mechanism. Reliability of services is an important aspect of the IoT network. Blockchain offers a solution that is safe and cost-effective in terms of reliability and cost, and blockchain technology has been introduced to manage the ranking of service providers sent by various IoT devices In this article, a system model can be proposed to protect IoT devices from unreliable services. The proposed system is written with the Atrium China blockchain platform and uses the POA consensus algorithm. This system allows IoT devices to know the rankings of service providers before requesting services. Smart contracts are introduced to store the rank of service providers in the Chinese bloc. IoT devices use the smart contract method to rank service providers. Performance analysis and experimental results show that the proposed model of IoT devices protects against unreliable services at a reasonable time and cost-effective [17].

Seller rating information from previous buyers gives new customers useful information when making a purchase decision. Thus, Bitcoin somehow obscures the relationship between buyer and seller with a layer of limited anonymity, thus preventing buyers from finding or confirming this information. While this level of anonymity is valued by society, as Bitcoin moves toward greater acceptance, buyers want to know more about who they trade with, and sellers who see their reputation as a strong selling point. , Are pressed to create more transparency. This paper proposes three different models by which a web credit or rating system can be implemented in connection with bitcoin transactions, and the advantages and disadvantages of each are explained. As a result, each of them poses challenges on both the technical and social fronts [18].

## 4 The proposed online rating system with blockchain

The proposed rating system is presented in this chapter. This system uses distributed ledger technology. Therefore, first, different types of distributed ledger technology are introduced along with their properties and it is argued why blockchain is the most suitable among them for the proposed system. in the proposed system, the Ethereum platform, which is a public permission network, is used. The main components of the proposed rating system and the smart contract written for it are presented in this chapter. To provide gas for the transactions, Ropsten and Rinkeby TestNets are used. All details are presented in this chapter.

## 4.1 Comparison of distributed ledger technologies

Considering the different features of each distributed ledger technology, one of them is chosen to be used in the online rating system. Different distributed ledger technologies are compared in Table 3.

**Table 3** Comparison of distributed ledger technologies.

| Feature | Hybrid | Blockchain | Directed acyclic graph (DAG) |
| --- | --- | --- | --- |
| Structure | Chain structure with a sequence of node hierarchies | A chain of information structures called blocks | A type of data structure called "guided diagram" which uses topological order methods for validating transactions. |
| Scalability | More than 40 million transactions per second | Depends on the blockchain platform | No limit on the number of transactions per second |
| Performance | Combining the features of both the blockchain technology and directed acyclic graph | Several blocks cannot be created simultaneously. Therefore, if there are too many transactions in the network, they should wait in a long queue for their turn to be inserted into new blocks because each block has a fixed size. | An effective and efficient tool for sending transactions throughout the network. Validations are not created with new blocks. New transactions are validated by older nodes. |
| Smart contract | Supported | Supported | Not supported |
| Free transaction | ---- | Cost (award) is received for each transaction validation by validators | Uses all the nodes willing to do a transaction for validating the transactions of the last two nodes. Therefore, there are no validators. |
| Programming development | BEXAM platform. This platform is not open-source yet | Solidity – Java – C# | --- |
| Privacy | ---- | There is privacy | There is privacy |
| Storage | ---- | Blocked structure. For each transaction, a block is added to the chain. Therefore, it requires lots of storage space over time. | There are no blocks in the structure and so it does not require a lot of storage. |
| Security | Security features similar to blockchain technology | Using encryption mechanism and hash for providing high security | This technology has lower security when the graph network is small, especially compared to blockchain. In that case, hackers can attack the graph because there are not many nodes to have good communication between them. |
| Validation | Using consensus algorithms like POS[3] DPOS[4] … | Anyone can become a validator by meeting certain criteria and receive awards by providing services to the network. Consensus algorithms for validation depend on the blockchain network type. | Relies on users (nodes) for validating transactions. |
| Offline capability | ---- | Transactions done during offline hours cannot be merged with the blockchain later. | A transaction can easily be reconnected to the main graph network, even after a long offline time period. |
| Cost | ---- | High cost | Transfers information faster and cheaper. |

---

[3] Proof of Stake
[4] Delegated Proof of Stake

After the comparison in Table 4, blockchain technology was chosen from distributed ledger technologies because blockchain technology supports smart contracts. Smart contracts guarantee decentralized communication and full trust between two parties, raters and products, in the online rating system without the need for intermediaries. The directed acyclic graph method does not support smart contracts, but the hybrid method is still new and no code or documentation has been published for it.

To present the workflow of the online rating system using blockchain technology with smart contracts, we propose a blockchain-based rating system to provide rating validation, rating data transparency, and raters' privacy in this system.

### 4.2 Online rating system with blockchain technology

The three indicators that the online rating system must have are transparency, privacy, and validation, which are expressed as follows:

1. Transparency: the final result must be transparent. This means that all raters must be sure their ratings have been counted.
2. Privacy: no one can know what rating the users have submitted.
3. Validation: it is signed using the user's private key (rating data), validators can verify it with the public key. Also, ratings cannot be manipulated.

Blockchain can provide these features. Blockchain technology allows the rating validation and the counting process to be done in a decentralized manner. The main advantage of using this method is that considering its decentralized nature, it can verify the accuracy of different stages of rating by the people involved in the network and all rating information is stored on the nodes participating in the network instead of servers. Therefore, if someone wants to manipulate the data, he must infiltrate all the nodes (computers) involved in this process and get access to a huge amount of unique data, which is considered practically impossible. Furthermore, blockchain technology allows a rater to easily check whether his or her rating has been submitted correctly. Also, in case there is any problem with the rating steps of the organization's products or services, it is automatically identified for the user. In addition to this, the transparency of blockchain makes monitoring and carrying out different steps of rating easier because independent people will be involved in the monitoring process. Also, in this method, there will be no need for spending extra resources or even the physical presence of employees or raters.

In addition to all these items, blockchain is encrypted in the rating system and supported by mathematical algorithms. Doing this helps with anonymity, hiding results, and carrying out calculations on encrypted data. These are the items that are not found in other blockchain-related systems due to their decentralized and open-source nature. Using this system, no one can ever make any changes to the rating results by manipulating data. Finally, the users can make sure their ratings and comments are submitted in the system after they are done with the rating process.

Another goal of creating this system is integrating the data used by the system which are complete, accurate, without paradoxes, and free of any type of logical errors. In fact, the system can prevent other users from accessing its information and only defined users will have access.

### 4.3 Main components of the blockchain rating system

The main components of the online rating system are presented in Figure 4. As seen in the figure, this system consists of three infrastructure, service, and business layers. In this system, we used the Ethereum blockchain platform, where for the blockchain network a public permission network is chosen, because users have access but cannot make ratings unless they submit a request in the system and get authenticated to participate in the online rating process of products. The most important feature of the Ethereum platform is its support for smart contracts. We can use this to write a rating contract using the Solidity language. Another feature of this platform is that we can use it to create a decentralized rating app. Also, all rating transactions are verified by the validators using license proof of authority consensus mechanisms. This way, raters' ratings are stored in the blockchain. Therefore, raters' privacy is maintained, it guarantees data transparency, and it creates rating data integrity.

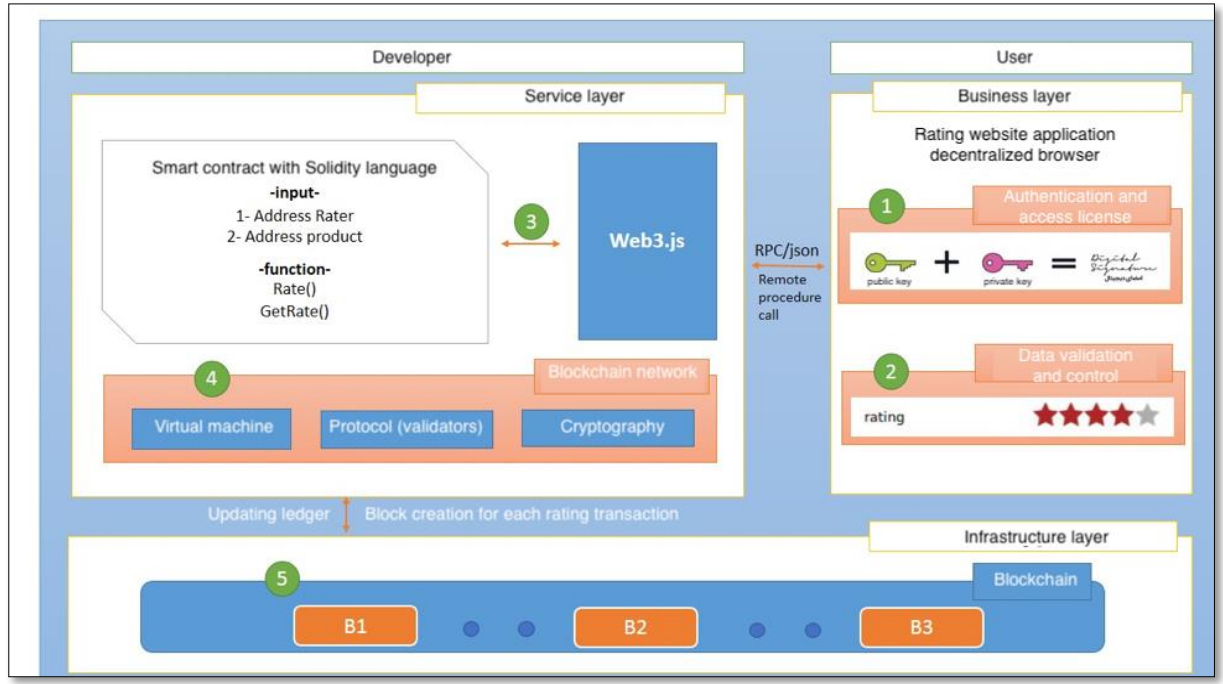

**Fig. 4.** Main components of the blockchain rating system.

**Business layer:** The business layer is the first and highest layer in the application. This layer provides the services and shows the content to the end-user through a graphical user interface. This layer cannot be accessed using any end-user device like desktops, laptops, smartphones, etc. Page contents and other display elements that are shown to the end-user need to be fetched by the web browser. This layer needs to communicate with other layers for displaying content. Only raters who have completed the approval process on the website that provides information can participate in the rating process.

This layer of the system describes procedures that interact with the user. When the user presses the rating button somewhere (web application, mobile, etc.), the smart contract is executed and a transaction is sent to the network. For each transaction, a block is added to the blockchain.

There are some tools in the business layer that the user can use to have a node in the Ethereum blockchain network. The Metamask tool (Ethereum wallet) allows users to set transaction fees dynamically and manage their custom tokens. Therefore, each user creates a user account using this tool, which includes a public key, a private key, and a digital signature as presented in figure 5 part 1. This tool is installed in one of the explorers as an extension and automatically connects to the Ethereum blockchain. Then, the user goes to the products page to select one of them and rate it. The next important component in this layer is control and validation. This component is for validating rating data received from the rater. For example, the rating is done by the rater to make sure whether the rating information is correct and accurate? Is the data valid?

**Service layer:** this layer deals with the developer and provides the needs of the higher "business layer". Several other components including smart contract, Truffle framework, and a component called transaction management, which is used to ensure the blockchain structure, is involved in this layer.
In this layer, JavaScript is used which has the following functionality:
1. It interacts with the web page HTML file. The rater calls the JavaScript component by submitting a rating. For each user transaction, a block is created and verified by the validators.
2. The smart contract generates the two ABI-Bytecode values. These values help us use the functionality and structure of the smart contract.
3. It is the communication of the smart contract with Web3.API so that we can communicate with the validators regarding transaction management and data validity.

**Infrastructure layer:** it contains the blockchain where each block includes the block details, a list of transactions, encryption of the last key, and the encryption of the block key. The rating is done by the user in the business layer and then, the transaction is transferred to the service and infrastructure layer using a remote procedure call (RPC).

When multiple transactions are registered on the Ethereum network, they are packed into one block and each block is linked to the block after that. This linked string of blocks that keeps the information on all transactions is the blockchain. To ensure that all nodes in the network have the same version of transaction data and to make sure no invalid data is recorded in the database, Ethereum uses an algorithm called Proof of Work (POW) or Proof of Authenticity (POA).

Workflow of the online rating system with blockchain architecture is presented in figure 5. In this figure, raters go to the web application and register. This way, each rater has a public and private key. Then, they are allowed to select one of the products and rate it. Of course, when the rater presses the rating button, the smart contract is executed and a transaction is added to the network. Then, this transaction is validated. Afterward, a block is added to the blockchain for these transactions. In the end, the blockchain and the result is updated completely.

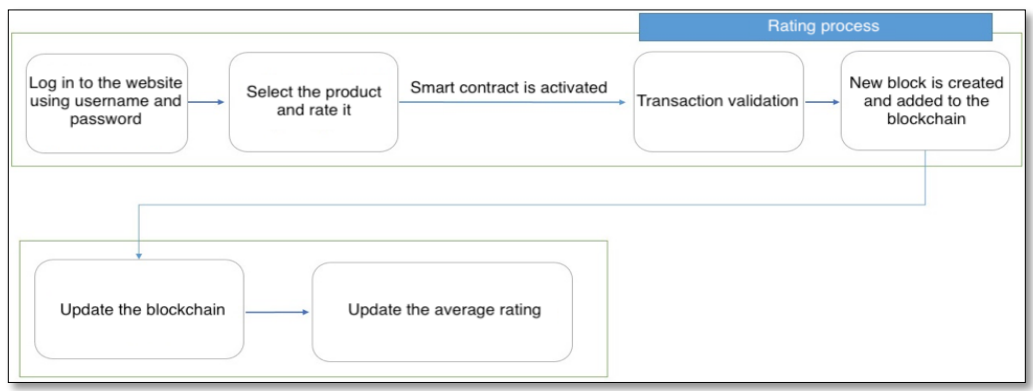

**Fig. 5.** Workflow of the online rating system using blockchain technology.

### 4.3.1 Executing the rating smart contract on the Ethereum blockchain platform

Ethereum has created a protocol for making decentralized applications. In a sense, Ethereum has created a fundamental layer for developers and a blockchain that is completely in the Touring programming language, which means if a system has unlimited resources, memory, computational power, and storage, they can execute infinite loops. Ethereum allows anyone to write their own rating smart contract and rating application. Also, they can set their rules for transaction ownership, format, or form, and transferring transactions. Since the rating smart contract uses blockchain, there is no need for an intermediary or supervisory institution for its execution. The rating smart contract is executed automatically on the blockchain network and has high security and speed and eliminates intermediation costs. Ethereum is a decentralized system and uses a peer to peer approach. Any transaction that takes place is supported by the users present in the networks and is not controlled by any centralized power. Rating smart contracts are developed based on a blockchain like Ethereum and prevent any type of forgery and cheating.

For a rating smart contract to be created and work correctly, they should operate in an appropriate and special environment. This environment should support public-key encryption to enable the raters to sign the transactions using their unique code. This is the system that almost all current cryptocurrencies use. The rating smart contract needs an open and decentralized database which gives the parties complete trust and is fully guaranteed. Rating smart contract activities and registrations are trackable and irreversible. Contracts written in such high-level languages are converted on the Ethereum Virtual Machine (EVM) and then registered on the Ethereum blockchain and executed.

The programming languages used for Ethereum Virtual Machine are the Solidity language, which is similar to the C and JavaScript language, and the Serpent language, which is similar to the Python language. Of course, the Solidity language is mostly recommended for writing different smart contracts. Other aforementioned languages are only used sometimes and in special cases.

### 4.3.2 Ethereum Virtual Machine

The Ethereum protocol has been designed such that it does more than peer to peer transaction processing. This program has been designed for executing complex codes and developers can easily perform tasks on it. Such that a system is needed for interpreting the commands and this is done using Ethereum virtual machines in Ethereum [19]. The smart contract is run on the Ethereum virtual machine. The Ethereum virtual machine checks all the information, including the rater's user account and the product chosen for rating, so that the transactions are stored correctly and logically. For these transactions, new blocks are created and saved in the blockchain.

### 4.3.3 Using TestNet blockchains in the rating system

Since we implement the online rating system on the public and permission blockchain network, some gas must be paid for each rating transaction. However, since this method is costly, the only solution to try the rating system is TestNet blockchain. In this solution, our network becomes private because TestNets can only be tried on a private system or small groups.

When we write an Ethereum virtual machine program on the Ethereum blockchain, we must pay gas for setting up and using that program. Gas in the Ethereum network is considered as the transaction fee. TestNets are extremely useful tools in developing Ethereum virtual machines. These TestNets make it easy to test Ethereum software and provide developers with a secure layer to test their program before running it on the main network. TestNets are almost like the main network in all aspects but the Ether on these networks is worthless. Public TestNets are available to anyone connected to the internet. All users can connect to these networks and connecting to them is even possible through wallets like Metamask [20].

One of these networks, Ropsten TestNet, was released in November 2016. Ether of this network can be mined exactly like the Ether of the main network. From these several Ethereum TestNets, Ropsten resembles the main network the most because its consensus mechanism is proof of work. Therefore, simulating transaction validations is closer to reality. In the Ropsten network, Ether can be mined or it can be received from the Faucet website, which is only there to give out trial Ethers.

To use the proposed online rating system based on the public permission network with the PoA mechanism, the Rinkeby TestNet can be used that supports the PoA mechanism. Therefore, in this type of network, any rater on the online rating system can send a registration request to the network so that the raters can grant rating permission. Other TestNets like Kavan are also very important for decentralized applications that need to interact with each other.

### 4.3.4 Rating Smart Contract

A smart contract resolves all concerns regarding any manipulation or cheating in the rating process by presenting a completely secure system. The ratings protected in the ledger must be decrypted and a lot of computational power is needed for accessing them. A big portion of the raters recognizes a bad and unreliable system as the reason for the lack of interest in rating. With smart contracts, participants can submit their ratings online and the rating system gets transformed immediately.

### A. Using the Solidity programming language in the rating system

This system is the code that enables the decentralized execution of contracts or applications. This language has been developed specifically for the Ethereum network and sometimes it is asked why the Ethereum virtual machine should have its specific language?

**Firstly:** we can convey difficult concepts specific to smart contracts with relatively easy commands in the Solidity language. During this short period of time, companies and developers have been able to use this programming language well and create several standards and libraries for developing smart contracts by sharing their achievements.

**Secondly:** any blockchain protocol, the Ethereum network for example, creates a decentralized and proprietary network on the internet and each of the users is a node on that network. The operation of decentralized networks is very different than the current conventional internet and it is also sometimes called the internet three. In the centralized rating system, all users' information is ultimately stored in centralized servers. However, Ethereum network users' information is stored on all network nodes, i.e. users, and there is no main server and this is why it is said to be decentralized. Anyway, these differences make the programming language different for smart contracts. For instance, there is no random function in the Solidity language because the Ethereum virtual machine requires complete certainty in smart contracts. If the smart contract acts uncertainly, since each node validates the block independently, the network literally stops working if the other nodes cannot reach an agreement. Another thing is that reading the Ethereum network and all other blockchain networks is free. However, transactions or registering and executing smart contracts on the Ethereum network have a cost in the form of Ethers.

Of course, simulated environments provide users with fake and free Ether cryptocurrency for testing the contract so that it can be tested several times for free and the possible required modifications can be made before it is registered and run on the main network. Smart contracts usually have a financial toll and making sure they work correctly is important. Also, it is absolutely impossible for anyone to change or maybe fix the contract after it has been registered and run on the main network. This multiplies the care put into testing the program. In the rating system smart contract, Solidity version 0.5.0 has been used. In figure 6 (line 1), the contract is called Transparent Rating (figure 6, line 3).

```
1  pragma solidity ^0.5.0;
2
3  contract transparentRating {
4      address private  owner;
5
6      modifier onlyOwner() {
7          require(msg.sender == owner);
8          _;
9      }
10
```

**Fig. 6.** Smart contract definition using the Solidity language

Figure 6 (line 6): Modifiers are extension functions that contain logical conditions. For example, there is a function for deactivating the active key and one modifier that needs to set the key status to enabled. If something is added to the function, it can only be called when the caller (msg.sender) is equal to the owner variable. At line 7, the required keyword expresses that everything inside the parenthesis must be true and otherwise Solidity will throw an error and stop the execution. The term "_;" at the end of the code expresses that after reading the modifier is finished, it will be the turn of the real function. The smart contract is a rating system between the product (service) user and the service or product provider. Here, two separate lists are needed to save the information of both parties.

```
16      struct RatingProducts {
17          string name;
18          uint rating;
19          uint no_raters;
20      }
21
22      struct Rater {
23          uint weight;
24          uint id;
25          bool rated;
26          uint rate;
27      }
28
29
30      mapping(address => Rater) public Raters;
31      mapping (address => RatingProducts) public ProductRate;
32
```

**Fig. 7.** Definition of the list of products and raters

Figure 7 shows the product provider at line 16 which includes name, current rating, and the number of raters that have rated that product. However, in line 22, participants in the system should be authenticated to participate in the rating process. Then, we specify whether the rater has already submitted their rating or not?

Figure 7 lines 30 and 31 show the type of address to the raters and providers and call it Raters Product Rate to save any reference address. This is a key/value data structure where the value can be obtained using the key and it is equivalent to the dictionary used in languages like Python and JavaScript.

```
43 -    function SetRate(uint newRatingGiven, address productAddress, address rater) public{
44          require(
45              msg.sender == owner,
46              "Only owner can give right to rate."
47          );
48          require( //
49              !Raters[rater].rated,
50              "The rater already rated."
51          );
52
53          uint oldRating;
54              uint newRating;
55                  uint no_raters;
56 -        if(newRatingGiven<=100){
57          oldRating = ProductRate[productAddress].rating;
58          no_raters = ProductRate[productAddress].no_raters;
59          newRating = oldRating + newRatingGiven;
60          // uint finalNewRating = getDivided(newRating, no_raters);
61          no_raters = no_raters +1 ;
62          ProductRate[productAddress].rating = getDivided(newRating, no_raters);
63          ProductRate[productAddress].no_raters = no_raters ;
64          }
65      }
66
```

**Fig. 8.** SetRate function

Figure 8: line 43 declares the SetRate function that gets the address of the rater along with the new rating value as input. When we call this function, we assign the new rating to the product that is being rated.

Figure 9: line 70 declares the GetRate function that gets the address of the input product to show the current result of each product. To test and execute the rating smart contract, the Remix environment is suitable for debugging the rating smart contract (Figure 10).

```
69
70 -    function GetRate (address productAddress) public returns (uint) {
71          uint Rating =  ProductRate[productAddress].rating;
72          return Rating;
73          }
```

**Fig. 9.** GetRate function

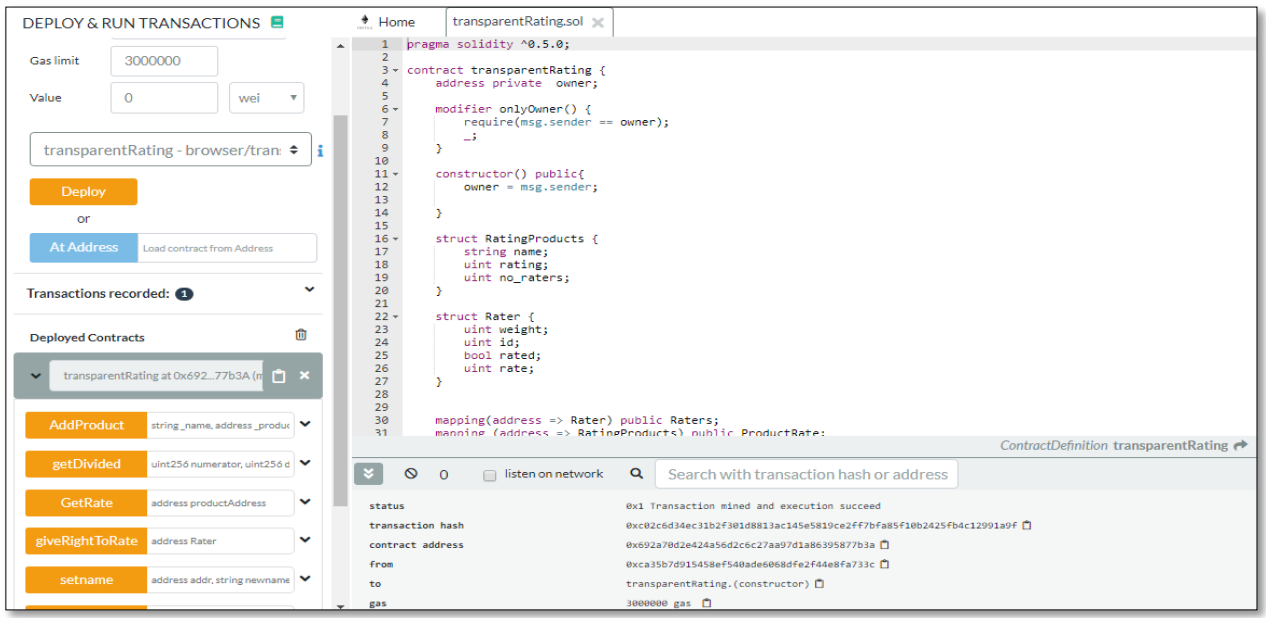

**Fig. 10.** Remix environment

### B. Executing the rating system smart contract

**Step one:** in Figure 10, we put the TransparentRating.sol file (smart contract) in IDE Remix. To execute the contract, we need an account and some Ether in the TestNet blockchain (Rinkeby or Ropsten network). Now that the contract is in IDE Remix, we have connected Metamask to the Ropsten TestNet that has an account and some Ether.

**Step two:** we click the Deploy (run) button. We choose Web3 Ropsten. Then, the Metamask account with balance is presented (Figure 11).

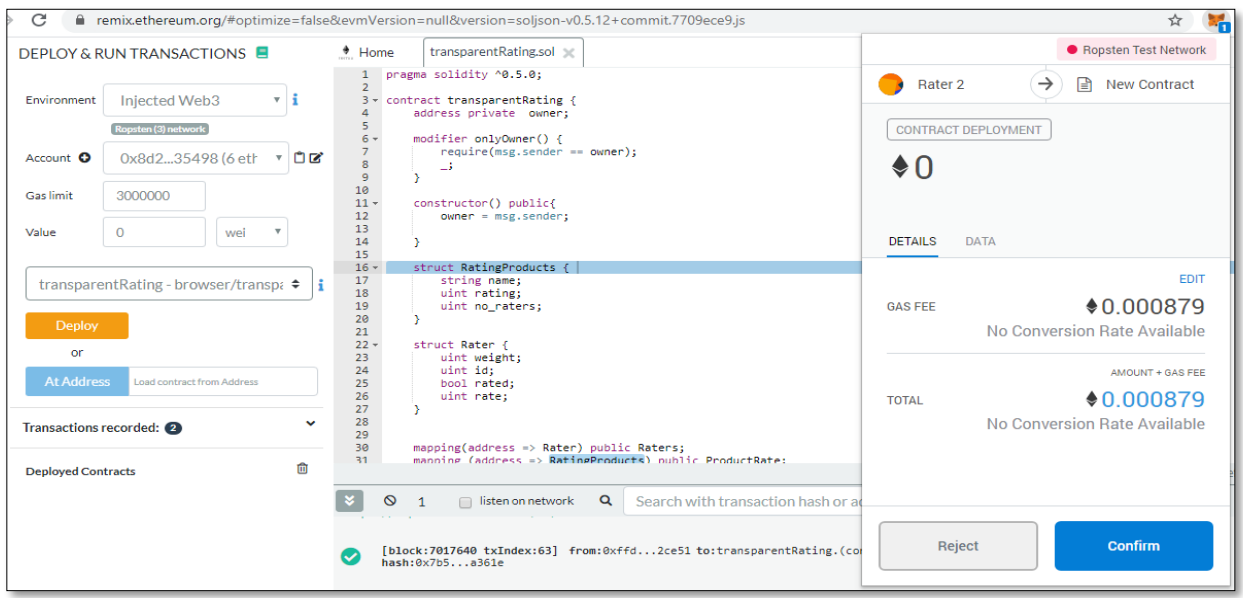

**Fig. 11.** Remix execution environment with the Metamask tool

To see the executed transaction and its details, we can take a look at the terminal of the Remix environment. The second way is to open the EtherScan to see the details as well (Figure 12). Therefore, the smart contract execution method has been done in the Ropsten public TestNet using Ethereum IDE Remix.

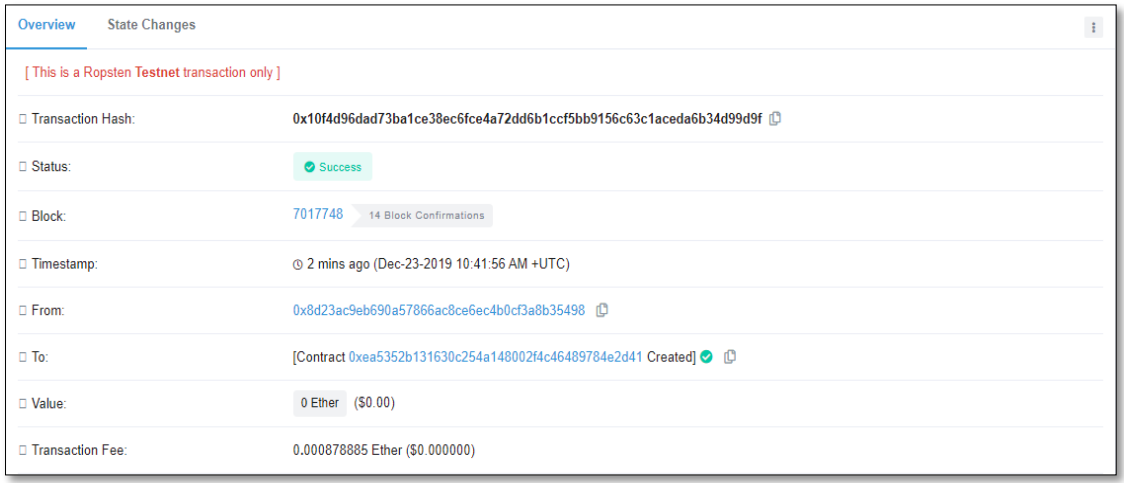

**Fig. 12.** Ether Scan website

Instead of connecting to public TestNets, as mentioned before, another method is connecting to the private or local blockchain network called Ganache. Ganache creates several Ethereum addresses where each address has an Ether value as presented in Figure 13. Since writing in the blockchain requires spending Ether, trial Ether should be used while testing rating smart contracts in TestNet blockchain. The only extra thing that should be done is to modify the Truffle settings file to conform to Ganache.

In Figure 13, Truffle uses a general ganache-cli version in its operational area and sets up a local network on port 8545 to show the 10 initial external accounts and its private keys for our convenience.

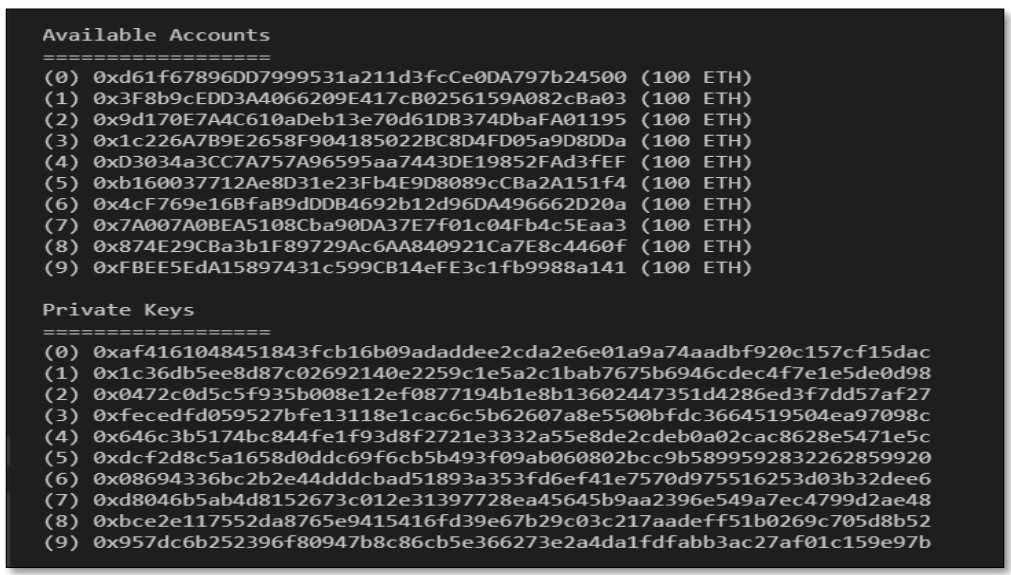

**Fig. 13.** Ten accounts with private keys

Therefore, to test the online rating system, a rating transaction should be sent to the blockchain network by the raters. To do this, one of the ten accounts presented to us in this tool can be used. Each account has its own private and public key. Therefore, we can use the public key to define this account in the system and carry out the registration process so that it has permission to register its rating on the system. By default, these accounts have worthless Ether or gas values so that infinite transactions can be done in the system and test transactions can be established in the system. To manage all the transactions on the blockchain network and to have an interactive console with the smart contract, a Truffle framework is required that has completely concentrated on coding and has no concern regarding blockchain network management. To install this framework, the following command needs to be entered into the terminal (Figure 14):

```
$ npm install -g truffle
```

**Fig. 14.** Installation method of Truffle by NPM

After installation, the following files and folders are created for writing the rating smart contract (Figure 15).

```
∨ contracts
    ♦ Migrations.sol
∨ migrations
    JS 1_initial_migration.js
∨ test
{} package-lock.json
JS truffle-config.js
```

**Fig. 15.** Created files and folders

Code of the contract prior to testing each unit. The following test equipment can be used (figure 16).

```javascript
1   const assert = require("assert");
2   const ganache = require("ganache-cli");
3   const Web3 = require("web3");
4
5   const web3 = new Web3(ganache.provider());
6   const { interface, bytecode } = require("../build/Token");
7
8   let accounts, token;
9   const INITIAL_SUPPLY = 100;
10
11  beforeEach(async () => {
12    accounts = await web3.eth.getAccounts();
13
14    token = await new web3.eth.Contract(JSON.parse(interface))
15      .deploy({
16        data: bytecode,
17        arguments: [INITIAL_SUPPLY]
18      })
19      .send({ from: accounts[0], gas: 1000000 });
20  });
```

**Fig. 16.** A code for contract test equipment.

During the tests, we used ganache to expand our contract to the local network. This way, to expand our contract with the public Ethereum network, we need a provider to connect to a real Ethereum node. We set up one but for simplicity, we use Infura. Infura is supported by ConsenSys and provides terminals that we can use to easily create a provider [21-24].

Infura provides terminals in the main networks and also in all TestNets. Therefore, when we

want to define a provider, it is as follows (Figure 17):

```
module.exports = {
    networks: {
        rinkeyb  : {
            provider: function(){
            return new HDWalletProvider(mnemonic,"https://Rinkeby.infura.io/v3/<INFURA_PROJECT_ID>")
            },
            network_id: 4
        }
    }
};
```

**Fig. 17.** Introduction of a public network like Infura

Therefore, when we use any type of public or private TestNet, rating will be done by raters for one or several transactions. A block is created where all the information is encrypted and this maintains the privacy of raters in the online rating system and it is for keeping the integrity of rating data and protecting them from any manipulation and abuse. In Figure 18, the Ganache graphical user interface for each block after being stored and validated is presented along with all the encrypted information.

```
CURRENT BLOCK    GAS PRICE       GAS LIMIT   NETWORK ID   RPC SERVER                MINING STATUS
14               20000000000     6721975     5777         HTTP://127.0.0.1:7545     AUTOMINING

← BACK    BLOCK 14

GAS USED    GAS LIMIT   MINED ON                BLOCK HASH
64156       6721975     2020-01-09 22:19:44     0×23d6e6a63af30285c88a89d940111807b476e5e1ea7b390f2ac8d83ac3299cab

TX HASH                                                                                                     CONTRACT CALL
0×bfe8d5cdd8152b2bd260bc42a45f9109da2750d6787361efe3ae3e466f89257a

FROM ADDRESS                                    TO CONTRACT ADDRESS                         GAS USED    VALUE
0×e2DdF51066dFE9158C402b341760D7868dff0cd7      0×920eFE7079F5f387Aefb8181144E4bB58ccA99aB  64156       0
```

**Fig. 18.** Saving a block in the blockchain with encrypted information in the Ganache tool

## 5  Evaluating the Performance

In this section, we review and evaluate the online rating system with blockchain according to the criteria and requirements of rating systems. In the case study part of this chapter, the rating smart contract functions and how much gas that must be paid for each function is studied. Then, two diagrams will be presented, one for when submitting ratings by the raters at different times varies depending on gas costs and then the diagram of the relationship between the number of raters and the time it takes for the transactions to be validated.

### 5.1 Criteria and requirements of the online rating system with blockchain technology

In an online rating system, users should be kept anonymous and data must be confidential. Blockchain technology can help in the online rating system. The online rating system proposed in this study has the following properties:

**Rater:** only eligible individuals can rate.

**Validation:** no rater can rate one product more than once.

**Rater privacy:** no one knows which rater has rated what product.

**Protecting rating data against manipulation:** no rater can change or manipulate the information.

**Participation:** establishing trust in the system leads to the higher participation of raters.

**Scalability:** this system has been designed for rating on a small scale.

**Gas cost:** each command has a fixed and predetermined gas cost. However, gas cost, Ethereum gas cost, is not constant. When a smart contract is called, we determine how much we intend to pat for each gas unit. In other words, the people who are the source of this computational power, i.e. Ethereum blockchain validators, decide whether they want to accept the proposed price and run the invoked contract. Since validators want to have more income in exchange for the processing power, they add to the Ethereum network, they prioritize transactions that pay more for the gas.

Another application for gas in the decentralized Ethereum network is a spam prevention [25], where a lot of transactions are sent in order to disrupt the network. The cost that must be paid for the gas for running the transaction and the smart contract prevents such attacks in the network.

Therefore, to create a trustworthy network with privacy where each product is registered with the correct rating by the raters which is done on the blockchain network, the following points can be used:

- In this study, we want to rate products and services. Considering the large number of products and services, we need a system for measuring transparency. Therefore, blockchain technology will have more transparency which makes different steps of a rating and presenting the rating results easier.
- If a product has a satisfactory rating result, it makes the rater choose this product and also the product owner will be happy because he has received a good rating.
- If the product does not have a satisfactory result, it prevents the rater from choosing it and he looks for another restaurant. On the other hand, competition arises such that the restaurant owner tries to improve its service quality.

### 5.2 Calculating the overall cost of each transaction

Each function in the smart contract burns some gas (Ether) for execution. We use Eq. (1). to calculate the overall cost of executing each function.

$$Ether = TXFee = GasUsed * Gas\Pr ice(Gwei) \quad (1)$$

Usually, as the price that the rater is willing to pay for gas goes higher, the value that validators receive from a transaction also increases. Therefore, it is more likely that validators will choose the intended transaction (Figure 19).

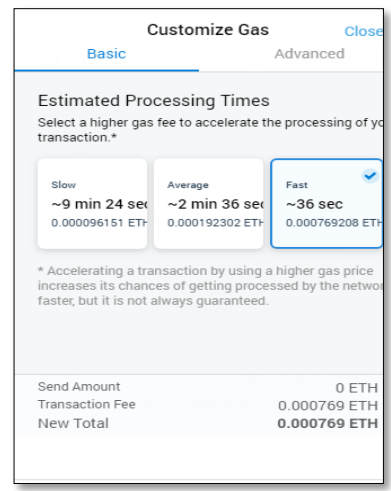

**Fig. 19.** Amount of gas and estimated time in the Ethereum wallet (Metamask tool).

Therefore, transaction validation speed increases as the gas capacity increases.

### 5.3 Case study

The proposed rating system has been tested on a website that displays restaurant information.

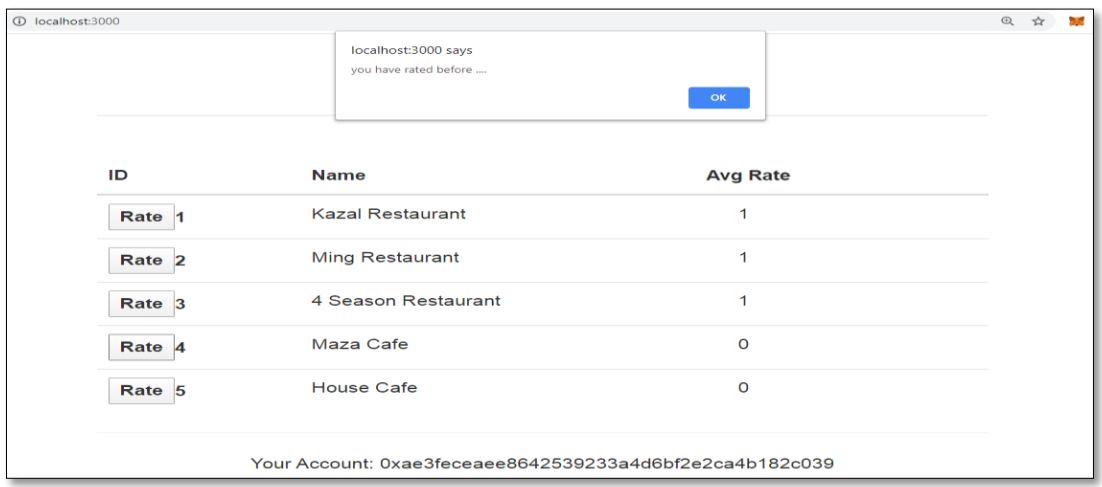

**Fig. 20.** Online rating system with blockchain technology.

In this system, eligible raters who have a license can declare their rating for the state of the restaurants. On this website, raters have their Ethereum accounts and users can open it in the web browser directly. Each expected features of the rating system with blockchain technology is discussed and studied here to find out if the feature is satisfied.

**Validation:** the blockchain system limits users from rating more than once. Such that each user signs its rating with his private key and submits it to the network using his public key. Figure 21 presents the error caused by a user rating twice.

**Fig. 21.** Error message for a rating submitted more than once

**Protecting rating data integrity against manipulation:** each time a rating is submitted, some information including rating submission time, smart contract address, and a list of transactions is encrypted and stored. Manipulating each block in the blockchain requires a lot of processing and during the limited available time, manipulation is impossible. A block along with the encrypted transaction stored inside it is presented in Figure 22. Also, Figure 23 shows a blockchain. In this figure, different blocks, their creation time, and the amount of gas spent on creating them is shown.

**Fig. 22.** Structure of a block created after a transaction is done in the Ganache tool.

**Fig. 23.** Different chained blocks in the Ganache tool.

**Smart contract with gas cost:** according to the smart contract presented in chapter four, three SetRate, GetRate, and GiveRightToRate functions exist in the proposed smart contract. The Metamask tool calculates the required gas for executing each one of these functions. It is worth noting that in the Ethereum TestNet, the cost of each gas unit (Ether) in equation 1 is assumed to be 1. Obviously, the real price of Ether replaces this value in the real world. Table 4 presents the cost of gas used for each function.

**Table 4** Smart contract functions and the gas cost for each one

| Rating smart contract functions | Gas required for execution |
|---|---|
| SetRate() | 51456 |
| GetRate() | 42689 |
| GiveRightToRate() | 47800 |

In a scenario, it is assumed that the rating system is being used for a Nov to Feb moons. The overall cost of the transactions of the rating system for 16 days has been calculated according to the gas price on that day and presented in Figure 24 and 25.

Since the price of Ether changes every day, to calculate the cost of using the proposed system, Ether price when the rating is being submitted should be considered.

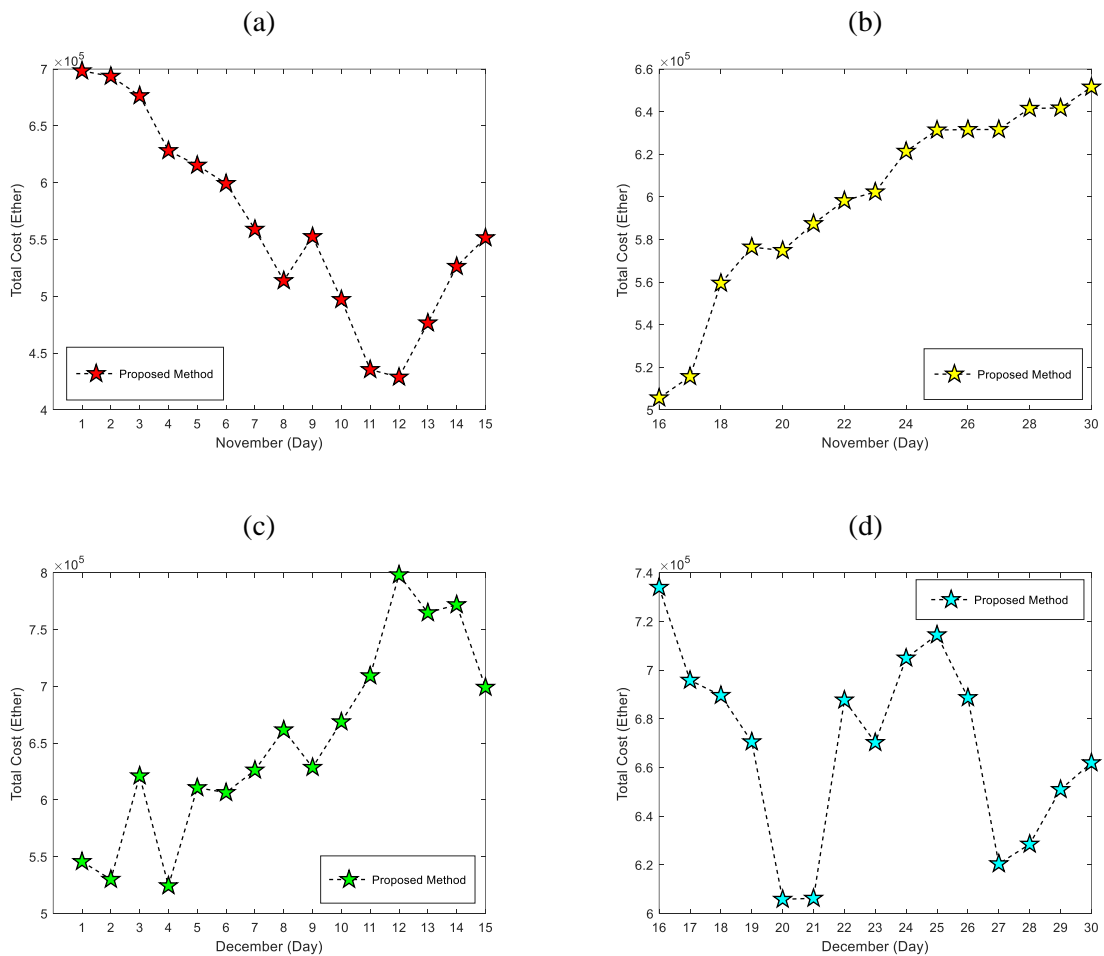

**Fig. 24.** Overall cost of transactions based on the transactions of each day (Nov and Dec)

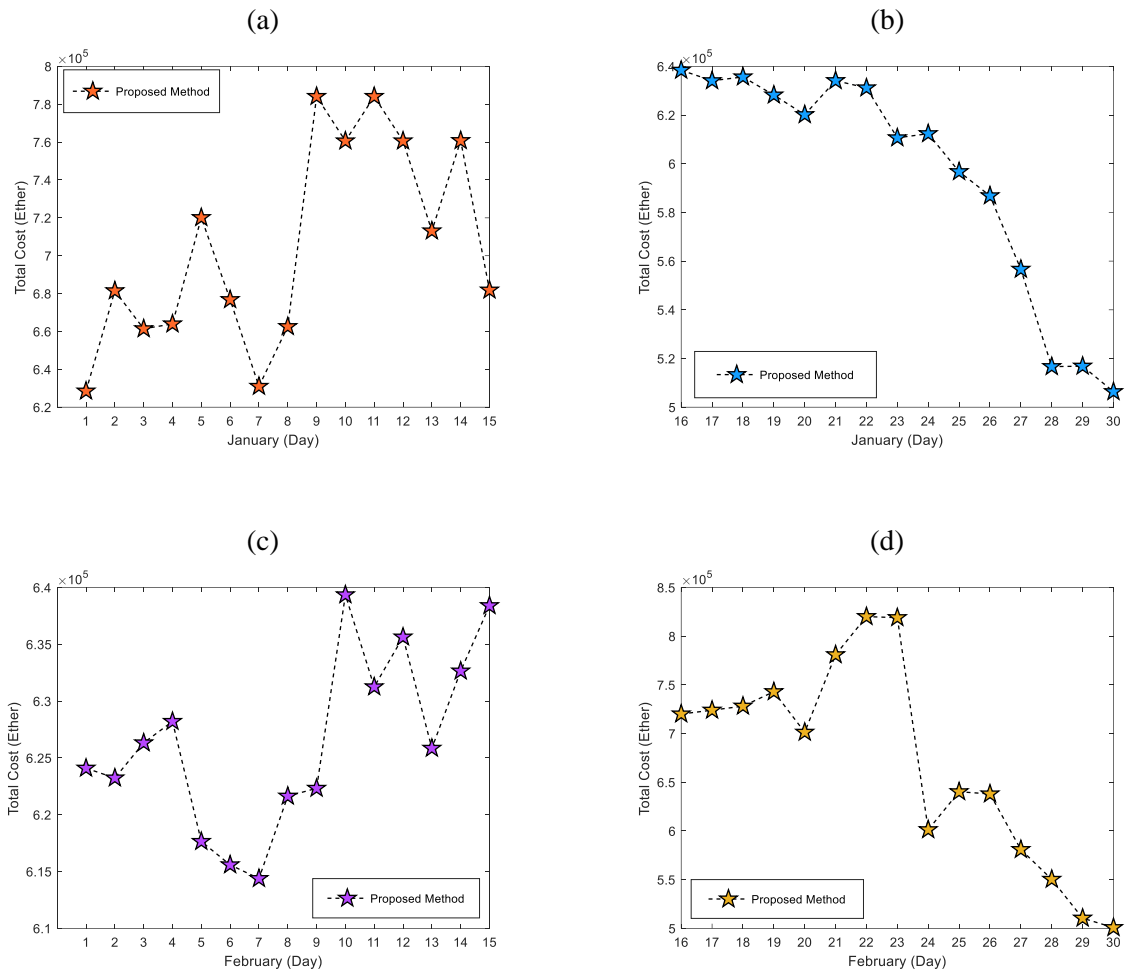

**Fig. 25.** Overall cost of transactions based on the transactions of each day (Jan and Feb)

### 5.4 Comparison of the current online rating system with the rating system with blockchain technology

Considering the weaknesses and challenges of the current online rating system, the rating system with blockchain technology solved the problems of the old system and provided different incentives for both the product and service raters and providers. In Table 5, the comparison of these two systems is presented:

**Table 5** Comparison of the current online rating system with the proposed system.

| Features | Online rating system with blockchain technology | Online rating system |
|---|---|---|
| System type | Decentralized system | Centralized system |
| Trust in results | Results are trusted | Rating of products is not trusted by the users and is not real |
| Abusing the results | Cannot be abused | Buying and selling ratings for products is possible |
| Information modification | No information can be altered or removed from the network | Information is controlled by the website manager |
| Decision-making | Since all information is accurately registered and has not been manipulated by any intermediary, reliable decision-making can be done | The right decision cannot be made about choosing the product |
| Raters | The product is selected in a short time and without any cost | More time and money spent on choosing products |
| Service providers | Since transparency is established and all the information of the raters is stored in the system, it creates a competitive or incentive system among all providers to present better services. | The system is not competitive and there is no control over the quality |
| Privacy | Because of using digital certificate technology for every rater, privacy is provided | Since the system is centralized, all the raters' information is kept by a manager and this eliminates privacy |
| Anonymity | Raters can submit their rating by creating a valid account and without any need for extra information and remain anonymous | Raters' information is known to the system manager |
| Paying the cost of rating | Rating is through gas (Ether) payment for the rating to be sent to the network | Raters submit their ratings without any cost |

## 6 Conclusion and Future work

In this research, an online rating system based on blockchain technology has been proposed. This system can be used in all product or service provider websites. Some of the advantages of using blockchain technology in the rating system are as follows. The database of such systems is decentralized and the data pertaining to ratings are distributed among the blocks and it is not possible for people to modify them, Blockchain provides privacy and creates trust in keeping rating data, this scheme has an online rating system and since blockchain technology prevents information modification, its results can be trusted, while using the rating smart contract, there are no intermediaries between the rater and the service provider in the rating system. The outcome of this research is the creation of a decentralized rating application using distributed ledger technologies, one of which is blockchain. Blockchain technology can prevent rating manipulation by protecting data integrity. This application has several features including the following: Making rating trustable and ratings citable, Rating is done online and exactly by the registered raters, A service or product provider can never manipulate rating results, Raters' personal information is kept confidential, no rater can rate one product more than once.

## Conflict of Interest

None.

## Reference

1. Heavin, C., & Power, D. J. (2018). Challenges for digital transformation–towards a conceptual decision support guide for managers. Journal of Decision Systems, 27(sup1), 38-45.
2. Zheng, Z., Xie, S., Dai, H., Chen, X., & Wang, H. (2017, June). An overview of blockchain technology: Architecture, consensus, and future trends. In 2017 IEEE international congress on big data (BigData congress) (pp. 557-564). IEEE.
3. Yelp, 2015. https://www.yelp.co.uk/
4. Kamran, M., Khan, H. U., Nisar, W., Farooq, M., & Rehman, S. U. (2020). Blockchain and Internet of Things: A bibliometric study. Computers & Electrical Engineering, 81, 106525.
5. Saad, A., & Park, S. Y. (2019, May). Decentralized Directed acyclic graph based DLT Network. In Proceedings of the International Conference on Omni-Layer Intelligent Systems (pp. 158-163).
6. Liu, X., Farahani, B., & Firouzi, F. (2020). Distributed Ledger Technology. In Intelligent Internet of Things (pp. 393-431). Springer, Cham.
7. Aung, Y. N., & Tantidham, T. (2019, February). Ethereum-based Emergency Service for Smart Home System: Smart Contract Implementation. In 2019 21st International Conference on Advanced Communication Technology (ICACT) (pp. 147-152). IEEE.
8. Zhang, Y., Kasahara, S., Shen, Y., Jiang, X., & Wan, J. (2018). Smart contract-based access control for the internet of things. IEEE Internet of Things Journal, 6(2), 1594-1605.
9. Lamba, A., Singh, S., Balvinder, S., Dutta, N., & Rela, S. (2017). Mitigating IoT Security and Privacy Challenges Using Distributed Ledger Based Blockchain (Dl-BC) Technology. International Journal for Technological Research In Engineering, 4(8).
10. Yuan, Y., & Wang, F. Y. (2018). Blockchain and cryptocurrencies: Model, techniques, and applications. IEEE Transactions on Systems, Man, and Cybernetics: Systems, 48(9), 1421-1428.
11. Schaufelbühl, A., Niya, S. R., Pelloni, L., Wullschleger, S., Bocek, T., Rajendran, L., & Stiller, B. (2019, May). EUREKA–a minimal operational prototype of a blockchain-based rating and publishing system. In 2019 IEEE International Conference on Blockchain and Cryptocurrency (ICBC) (pp. 13-14). IEEE.
12. Arif, Y. M., Nurhayati, H., Harini, S., Nugroho, S. M. S., & Hariadi, M. (2020, February). Decentralized Tourism Destinations Rating System Using 6AsTD Framework and Blockchain. In 2020 International Conference on Smart Technology and Applications (ICoSTA) (pp. 1-6). IEEE.
13. Yang, C. N., Chen, Y. C., Chen, S. Y., & Wu, S. Y. (2019, March). A reliable E-commerce business model using blockchain based product grading system. In 2019 IEEE 4th International Conference on Big Data Analytics (ICBDA) (pp. 341-344). IEEE.
14. Yang, J., Paudel, A., & Gooi, H. B. (2019, August). Blockchain Framework for Peer-to-Peer Energy Trading with Credit Rating. In 2019 IEEE Power & Energy Society General Meeting (PESGM) (pp. 1-5). IEEE.
15. Wu, H. T., Su, Y. J., & Hu, W. C. (2017, August). A Study on Blockchain-Based Circular Economy Credit Rating System. In International Conference on Security with Intelligent Computing and Big-data Services (pp. 339-343). Springer, Cham.
16. Mahajan, Y., & Srivastava, S. (2018, September). Holistic Credit Rating System for Online Microlending Platforms with Blockchain Technology. In International Symposium on Security in Computing and Communication (pp. 605-616). Springer, Singapore.
17. Iftikhar, M. S., Khan, Z. A., Noshad, Z., Khalid, A., & Javaid, N. (2019, November). Reliable Services from Service Providers Based on the Ratings of IoT Devices Using Blockchain. In 2019 Sixth HCT Information Technology Trends (ITT) (pp. 73-78). IEEE.
18. Vandervort, D. (2014, March). Challenges and opportunities associated with a bitcoin-based transaction rating system. In International Conference on Financial Cryptography and Data Security (pp. 33-42). Springer, Berlin, Heidelberg.
19. Park, J. S., Youn, T. Y., Kim, H. B., Rhee, K. H., & Shin, S. U. (2018). Smart contract-based review

system for an IoT data marketplace. Sensors, 18(10), 3577.
20. Fusco, F., Lunesu, M. I., Pani, F. E., & Pinna, A. (2018). Crypto-voting, a Blockchain based e-Voting System. In KMIS (pp. 221-225).
21. Aggarwal, S., Chaudhary, R., Aujla, G. S., Kumar, N., Choo, K. K. R., & Zomaya, A. Y. (2019). Blockchain for smart communities: Applications, challenges and opportunities. Journal of Network and Computer Applications.
22. Sengupta, J., Ruj, S., & Bit, S. D. (2019). A comprehensive survey on attacks, security issues and blockchain solutions for IoT and IIoT. Journal of Network and Computer Applications, 102481.
23. Wang, L., Shen, X., Li, J., Shao, J., & Yang, Y. (2019). Cryptographic primitives in blockchains. Journal of Network and Computer Applications, 127, 43-58.
24. Makhdoom, I., Abolhasan, M., Abbas, H., & Ni, W. (2019). Blockchain's adoption in IoT: The challenges, and a way forward. Journal of Network and Computer Applications, 125, 251-279.
25. Yavuz, E., Koç, A. K., Çabuk, U. C., & Dalkılıç, G. (2018, March). Towards secure e-voting using ethereum blockchain. In 2018 6th International Symposium on Digital Forensic and Security (ISDFS) (pp. 1-7). IEEE.